\documentclass[twocolumn,preprintnumbers,superscriptaddress,amsmath,amssymb,longbibliography]{revtex4-2}

\usepackage{graphicx}
\usepackage{dcolumn}
\usepackage{float} 
\usepackage{bbm}
\usepackage[dvipsnames]{xcolor}
\usepackage[colorlinks,linkcolor=blue,urlcolor=blue,citecolor=blue,anchorcolor=blue]{hyperref}
\usepackage{setspace}
\usepackage{xpatch} 
\usepackage{soul}

\usepackage[dvipsnames]{xcolor}

\begin{document}
\title{Measurement-based feedback as a flexible tool to control non-Poissonian photon statistics}

\author{Ming Li}

\affiliation{Shenzhen International Quantum Academy, Shenzhen 518048, China}

\affiliation{Shenzhen Institute for Quantum Science and Engineering, Southern University of Science and Technology, Shenzhen 518055, China}

\author{JunYan Luo}
\affiliation{Department of Physics, Zhejiang University of Science and Technology, Hangzhou 310023, China}

\author{Georg Engelhardt}
\email{georg-engelhardt-research@outlook.com}
\affiliation{Shenzhen International Quantum Academy, Shenzhen 518048, China}

\date{\today}

\pacs{
  }

\begin{abstract}
Non-Poissonian photon statistics are usually traced to nonlinear interactions, nonclassical states, or engineered reservoirs. Here we show that measurement-based feedback can be alternatively deployed to  control the Poissonian character of the measurement statistics in an externally-driven linear cavity. Each photon detection triggers a weak displacement operation whose phase determines how the feedback interferes with the coherent cavity field, resulting in  a significant cavity response.  Full counting statistics and two-time photon correlations show that the feedback phase controls the sign and strength of the induced correlations, continuously tuning the emission statistics between super- and sub-Poissonian regimes. At generic phases, the Mandel parameter responds linearly to the feedback amplitude, whereas at specific phases, the leading response becomes quadratic. The same phase dependence is also found in higher-order  cumulants of the emission statistics. These results establish measurement-based feedback as a route to controlling photon statistics in otherwise linear driven-dissipative systems.
\end{abstract}

\maketitle

\allowdisplaybreaks

\section{Introduction}
Photon statistics lie at the heart of quantum optics \cite{Glauber1963,Mandel1995}, and controlling them is essential for engineering fluctuations and correlations in emitted light \cite{Wiseman2009}, underpinning applications in quantum sensing \cite{Degen2017}, quantum communication \cite{Pirandola2020}, and quantum information processing \cite{Flamini2019}. Photon-counting records characterize these statistics at the level of individual emission events. Stationary, independent, and memoryless emission gives Poissonian counting statistics, whereas super- and sub-Poissonian statistics signal enhanced or suppressed fluctuations relative to this benchmark \cite{Mandel1979,Davidovich1996,Mandel1995}. 

A coherently driven linear cavity realizes this Poissonian reference in the absence of thermal noise and nonlinear interactions \cite{Glauber1963States,Mandel1995,Walls2008}. Non-Poissonian photon statistics have been extensively studied in quantum optics \cite{Paul1982,Davidovich1996} and  microwave quantum systems \cite{Goetz2017,Brange2019,Nesterov2020,Simoneau2022, Beaudoin2024}. They can arise through several established mechanisms, including nonlinear interactions
\cite{Kimble1977,Rempe1991,Birnbaum2005,Faraon2008},
thermal fluctuations \cite{Goetz2017,Brange2019}, 
parametric driving \cite{Portugal2023, Lemonde2014,Simoneau2022}, and postprocessing~\cite{Walker1986,Nunn2023}.

Here, we investigate a coherently driven linear cavity subject to click-triggered displacement feedback for its potential to generate non-Poissonian photon-counting statistics. Each detected photon triggers a coherent shift operation, which then profoundly reshapes the statistics of the subsequent emissions. The model and protocol are platform-independent, while superconducting microwave circuits provide a natural experimental motivation because they combine tunable microwave-photon control
\cite{Yin2013,Blais2021},
single-photon detection
\cite{Inomata2016,Besse2018,Kono2018,Lescanne2020},
phase-controlled pulses, and real-time measurement feedback
\cite{Vijay2012,Riste2012,Riste2013,Andersen2019,Minev2019}.
Existing superconducting-feedback experiments have mainly focused on state preparation, stabilization, and error correction
\cite{Vijay2012,Riste2012,Riste2013,Andersen2019,Minev2019},
leaving open how measurement-triggered control modifies the photon-emission statistics.  

In the short-delay limit, such conditional control can be described by a Markovian feedback master equation \cite{Wiseman1994, Liu2017}.
Full counting statistics, characterizing the accumulative probability distribution  of detected photons, provides a  complementing framework for quantifying how this causal structure reshapes the emission process~\cite{Levitov1996, Schoenhammer2007}.
In microwave cavities, this framework has been used to analyze emitted-photon distributions, waiting-time statistics, and long-time counting fluctuations
\cite{Brange2019,Nesterov2020,Portugal2023,Xu2023,Tang2026,Wu2026}, and the quantum information of dispersive readout~\cite{2s1m-y9bd}. Full-counting statistics have also deployed to charactize energy transport in Floquet theory~\cite{Engelhardt2024a,Engelhardt2024} and measurement statistics in spectroscopy~\cite{Engelhardt2025,Engelhardt2026}.
More recently, it has been extended to jump-based feedback protocols involving stored detection records
\cite{Rosal2026_2}.

Using this framework, we show that the feedback phase continuously controls the long-time photon statistics, producing both super-Poissonian and sub-Poissonian regimes, as characterized by the celebrated Mandel parameter. We identify the excess photon number following a single feedback operation as the microscopic origin of this behavior. Near  special feedback phases, interference suppresses the leading feedback contribution and changes the weak-feedback scaling of the Mandel parameter from linear to quadratic. The same mechanism extends to higher-order  cumulants. These results demonstrate that measurement-conditioned dynamics can engineer photon-counting statistics without intrinsic nonlinear interactions.

The remainder of this paper is organized as follows. In Sec.~\ref{Sec:ModelFCS}, we introduce the click-triggered displacement-feedback model and formulate the counting-field-resolved master equation. In Sec.~\ref{Sec:LongFCS}, we characterize the feedback-induced modification of the stationary cavity field and photon occupation. In Sec.~\ref{Sec:LongFCS}, we examine the two-time photon correlations and show how the feedback phase controls the transition between super- and sub-Poissonian counting. In Sec.~\ref{Sec:LongFCS}, we analyze the phase dependence and weak-feedback scaling of the Mandel parameter, including the change from linear to quadratic scaling at the quadrature phases and its dependence on detuning. In Sec.~\ref{Sec:LongFCS}, we extend the analysis to higher-order  cumulants. Finally, Sec.~\ref{Sec:Conclusion} summarizes our main results and conclusions.

\section{Feedback System and full-counting statistics}
\label{Sec:ModelFCS}

\begin{figure}[t]
\center
\includegraphics[width=0.95\columnwidth]{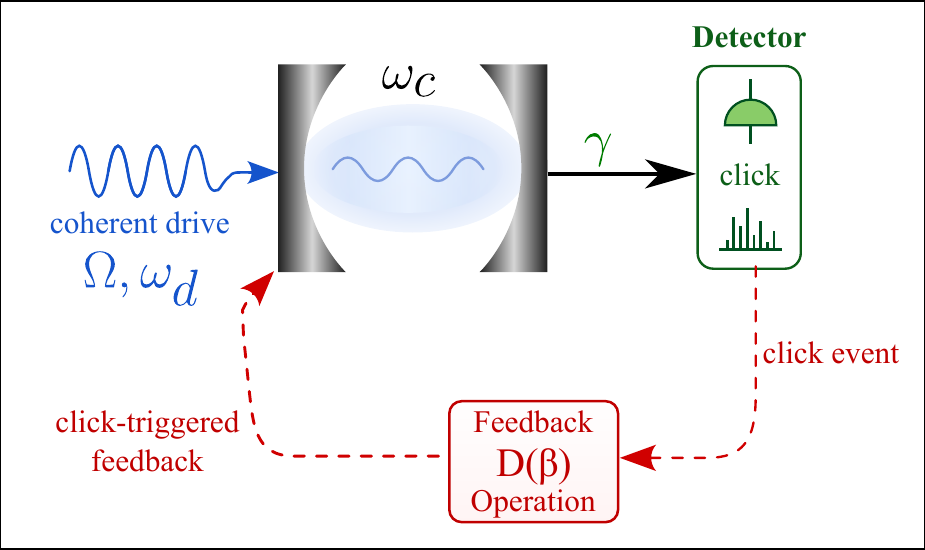}
\caption{Schematic of a coherently driven cavity under measurement-based feedback.
Photons emitted from the cavity are monitored by a detector, and each
detection event triggers a feedback displacement $\hat{D}(\beta)$ of the
intracavity field.
}
\label{Fig:Setup}
\end{figure}

We consider a coherently driven single-mode cavity coupled to a monitored photonic output channel depicted in Fig.~\ref{Fig:Setup}. In the rotating frame of the driving field, the  Hamiltonian describing the system and the output channels  is
\begin{equation}
\hat H_{\mathrm{tot}}
=\!
\hat H_S
+
i\sqrt{\frac{\gamma}{dt} }
\left(
\hat b_t^\dagger \hat a
-
\hat a^\dagger \hat b_t
\right).
\label{eq:microscopic_model}
\end{equation}
Here, 
\begin{equation}
	\hat H_S=\Delta \hat a^\dagger \hat a+\Omega \hat a^\dagger+\Omega^*\hat a
	\label{eq:systemHamiltonian}
\end{equation}
denotes the system Hamiltonian, where $\hat a$ annihilates a cavity photon. The cavity is thereby subject to a coherent driving field with Rabi frequency $\Omega$. The output channel is quantized by the photonic operators $\hat b_t$, which are labeled by the time $t$ at which they interact with the system~\cite{Gardiner2004,Wiseman2009,Clerk2010}. The corresponding cavity-loss rate is denoted by  $\gamma$. Prior to the interaction with the cavity, the photonic mode $\hat b_t$  is in the vacuum state $|0\rangle_t$.

The interaction is determined by the  short-time  propagator $\hat V_t(dt)=\exp[-i\hat H_Sdt+\sqrt{\gamma dt}(\hat a\hat b_t^\dagger-\hat a^\dagger\hat b_t)]$. 
After the interaction, the photon number of the outgoing temporal mode $\hat b_t$ is measured. In the limit $dt\rightarrow0$, the relevant outcomes are $m=0,1$, and the associated cavity measurement operators are $\hat M_m={}_t\langle m|\hat V_t(dt)|0\rangle_t$. In particular, $\hat M_1=\sqrt{\gamma dt}\,\hat a+O(dt^{3/2})$, such that a click, i.e., a measurement event, heralds the detection of one photon in the output channel. Each click triggers the instantaneous intracavity displacement $\hat U=\hat D(\beta)$, with $\hat D(\beta)=\exp(\beta\hat a^\dagger-\beta^*\hat a)$, while no feedback operation is applied for $m=0$. 

This time-resolved measurement protocol is used to construct the full counting statistics in terms of a generalized density matrix $\hat\rho_\chi(t)$: Before averaging over the measurement record, we associate a factor $e^{i\chi m}$ with the outcome $m$, giving
\begin{equation}
\hat\rho_\chi(t+dt)
=
\sum_{m=0,1}
e^{i\chi m}
\hat U^m \hat M_m
\hat\rho_\chi(t)
\hat M_m^\dagger \hat U^{\dagger m}.
\label{eq:microscopic_counting_map}
\end{equation}
Thus, each detected photon both increments the counting record and triggers a feedback displacement.

Taking the continuous-time limit of Eq.~\eqref{eq:microscopic_counting_map}
gives the dynamics of the generalized reduced density matrix of the resonator system $\hat\rho_\chi$,
\begin{equation}
\dot{\hat\rho}_\chi
=
\left[
\mathcal L
+
\left(e^{i\chi}-1\right)\mathcal J
\right]
\hat\rho_\chi
\equiv
\mathcal L_\chi\hat\rho_\chi .
\label{eq:tilted_liouvillian}
\end{equation}
The superoperators $\mathcal L$ and $\mathcal J$ are defined, for an
arbitrary system operator $\hat X$, by
$\mathcal J\hat X
=\gamma\hat U\hat a\hat X\hat a^\dagger\hat U^\dagger$
and
$\mathcal L\hat X
=-i[\hat H_S,\hat X]
+\gamma\mathcal D[\hat U\hat a]\hat X$,
where
$\mathcal D[\hat A]\hat X
=\hat A\hat X\hat A^\dagger
-\frac{1}{2}\{\hat A^\dagger\hat A,\hat X\}$ is the common  dissipator.
At \(\chi=0\), \(\hat\rho_{\chi=0}=\hat\rho\), and the tilted dynamics reduces to the common feedback master equation. The counting field therefore tracks the photons detected in the monitored output channel, which are precisely the events that trigger the feedback displacement. We consider the ideal zero-delay feedback limit and an instantaneous displacement following each photon detection. A detailed derivation of Eq.~\eqref{eq:tilted_liouvillian} is given in Appendix~\ref{app:microscopic_derivation}.

For a photon-counting record \(\{m_t\}\), the accumulated number of detected photons is \(N(\tau)=\sum_{t<\tau}m_t\). Its statistics can be expressed in terms of the moment-generating function
\(M(\chi,\tau)=\langle e^{i\chi N(\tau)}\rangle =\mathrm{Tr}[\hat\rho_\chi(t)] \), or, alternatively, the
 cumulant-generating function \(K(\chi,t)=\ln M(\chi,t)\)~\cite{2s1m-y9bd}. In the  long-time limit,
the cumulant rates are defined as
\begin{equation}
\kappa_m
=
\lim_{t\to\infty}
\left.
\frac{1}{t}
\frac{\partial^m K(\chi,t)}
{\partial(i\chi)^m}
\right|_{\chi=0}.
\label{eq:cumulant_rates}
\end{equation}
The first two cumulant rates give the stationary emission rate $\kappa_1$ and the
corresponding measurement noise $\kappa_2$, respectively, while higher cumulants characterize the
non-Gaussian counting fluctuations. We quantify the deviations from
Poissonian statistics by the Mandel parameter
\begin{equation}
Q
=
\frac{\kappa_2-\kappa_1}{\kappa_1}.
\label{eq:mandel_parameter}
\end{equation}
Thus, $Q>0$ and $Q<0$ indicate  super- and sub-Poissonian counting,
respectively, which correspond to bunching or anti-bunching of the photon detection events.

\section{Feedback control of cavity response and photon-counting statistics}
\label{Sec:LongFCS}

Having established the feedback model and the counting-field formulation, we now turn to the physical consequences of the click-triggered displacement. We focus on how the feedback phase modifies the stationary cavity response, two-time photon correlations, and the counting fluctuations, with particular attention to their phase dependence and weak-feedback scaling. We also examine how these features appear in higher-order  cumulants.

\subsection{Stationary cavity response}
\label{sec:stationary_response}

We first examine how click-triggered feedback modifies the stationary intracavity field.  For $\beta=0$, the stationary state of the cavity  is a coherent state 
\begin{equation}
	\rho_{\rm ss} = \left|\alpha_0 \right> \left< \alpha_0\right|
	\label{eq:stationaryState}
\end{equation}
with  $\alpha_0 = -i\Omega/(\gamma/2+i\Delta) $ and $\hat a \left|\alpha_0 \right> =\alpha_0 \left|\alpha_0 \right>$.   For finite but small $\beta$, a mean-field analysis reveals that the feedback-induced shift of the stationary amplitude $ \alpha_{\rm ss} = \langle \hat a\rangle_{\rm ss}$ for small $\beta$ becomes
\begin{eqnarray}
    \delta\alpha&=&  \alpha_{\rm ss}-\alpha_0  \nonumber\\
    &=& 
    \frac{\gamma\beta}
    {\gamma/2+i\Delta}
    |\alpha_0|^2
    +
    O(|\beta|^2),
    \label{eq:meanField}
\end{eqnarray}
as shown in Appendix~\ref{app:mean_field}.  Figure~\ref{Fig:Intracavity}(a) depicts the real and imaginary parts of $\delta\alpha/|\alpha_0|$ as the feedback phase $\Delta\Phi = \arg \beta$ is varied. 
For the resonant parameters considered here, $\alpha_0$ is real in our phase convention, so the two components have a simple phase-space interpretation. 
$\mathrm{Re}\,\delta\alpha$ describes the response in phase with  the stationary field, while $\mathrm{Im}\,\delta\alpha$ describes the response phase shifted by $\pi/2$.

The feedback phase controls the direction of this stationary displacement. 
Around $\Delta\Phi=0$ and $\pi$, the response is predominantly parallel to the original cavity field and respectively enhances or suppresses its magnitude. 
Near $\Delta\Phi=\pm\pi/2$, the feedback displacement is nearly orthogonal to the  stationary field without feedback, so its parallel component vanishes to leading order while the orthogonal response remains finite. 
The mean-field theory agrees well with the exact stationary-state results in the weak-feedback regime considered here.
A weak displacement applied after each detection event can nevertheless produce a significant stationary-field response through the compounding action of the feedback.

The corresponding photon-number dependence is shown in Fig.~\ref{Fig:Intracavity}(b). 
The mean intracavity occupation $\bar n=\langle \hat a^\dagger\hat a\rangle_{\rm ss}$ varies considerably with $\Delta\Phi$, reflecting the phase-dependent displacement of the stationary field. 
In contrast, the normalized intracavity photon-number variance $V_{n,\mathrm{cav}}=\mathrm{Var}(n)/\bar n$ changes only weakly over the same phase cycle. 
The feedback therefore shifts the center of the stationary photon-number distribution more strongly than it modifies its relative width.

The same stationary occupation directly determines the first  cumulant. 
Since the feedback displacement following a detected photon is unitary, the stationary photon-emission rate satisfies
\begin{equation}
    \kappa_1
    =\mathrm{Tr}\!\left[\mathcal J\rho_{\rm ss}\right]
    =\gamma\bar n .
\end{equation}
The phase dependence of the mean output photon flux therefore follows directly from that of the stationary intracavity occupation in Fig.~\ref{Fig:Intracavity}(b). 
The first cumulant, however, characterizes only the mean emission rate and contains no information about correlations between distinct emission events. These correlations become visible in the counting fluctuations, which we examine next.

\begin{figure}[t]
\center
\includegraphics[width=0.95\columnwidth]{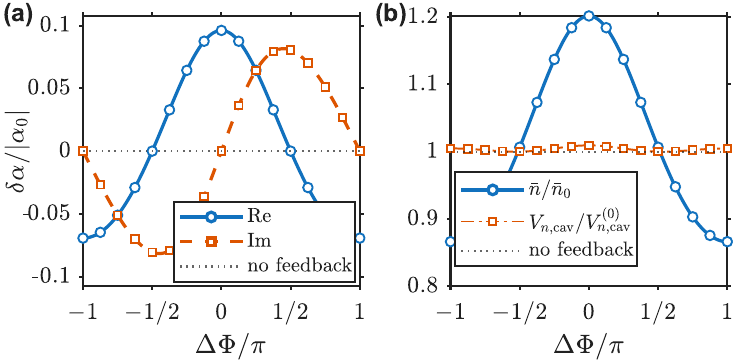}
\caption{
Stationary cavity response to click-triggered feedback.
(a) Normalized feedback-induced shift of the stationary cavity amplitude,
$\delta\alpha/|\alpha_0|$.
Symbols show the exact stationary-state results, while lines show the mean-field prediction.
(b) Mean intracavity photon number $\bar n/\bar n_0$ and normalized photon-number variance
$V_{n,\mathrm{cav}}/V_{n,\mathrm{cav}}^{(0)}$
as functions of the feedback phase $\Delta\Phi$.
For $\bar n/\bar n_0$, symbols denote exact results and the solid line is the mean-field prediction;
the dash-dotted line connecting the variance data is a guide to the eye.
The dotted horizontal lines indicate the corresponding no-feedback references.
}
\label{Fig:Intracavity}
\end{figure}

\subsection{Phase-controlled photon-counting fluctuations}
\label{sec:counting_fluctuations}

As shown in Sec.~\ref{sec:stationary_response}, the normalized intracavity photon-number variance changes only weakly under feedback. 
This single-time quantity, however, does not determine the long-time counting statistics, which are sensitive to temporal correlations between photon-detection events.

We consider the cavity state conditioned on a photon detection at $t=0$, which is given by 
$\rho_c(0^+)=\mathcal J\rho_{\rm ss}/\kappa_1$. 
The system then evolves under the full physical Liouvillian $\mathcal L$, which includes all possible intermediate emission events and their associated feedback operations. 
The normalized two-time detection correlation is ~\cite{Landi2024}
\begin{equation}
g^{(2)}(\tau)
=
\frac{
{\rm Tr}
\left[
\mathcal J
e^{\mathcal L\tau}
\mathcal J
\rho_{\rm ss}
\right]
}
{\kappa_1^2}.
\label{eq:g2_feedback}
\end{equation}
The long-time cumulant hierarchy gives the exact relation
\begin{equation}
Q
=
2\kappa_1
\int_0^\infty
\left[
g^{(2)}(\tau)-1
\right]
d\tau .
\label{eq:Q_exact}
\end{equation}
Thus, the deviation from Poissonian statistics is determined by the integrated correlation response rather than by \(g^{(2)}(\tau)\) at a single time.
A positive integrated correlation area gives $Q>0$ and hence super-Poissonian statistics, whereas a negative area gives $Q<0$ and sub-Poissonian statistics.

\begin{figure}[t]
\center
\includegraphics[width=0.95\columnwidth]{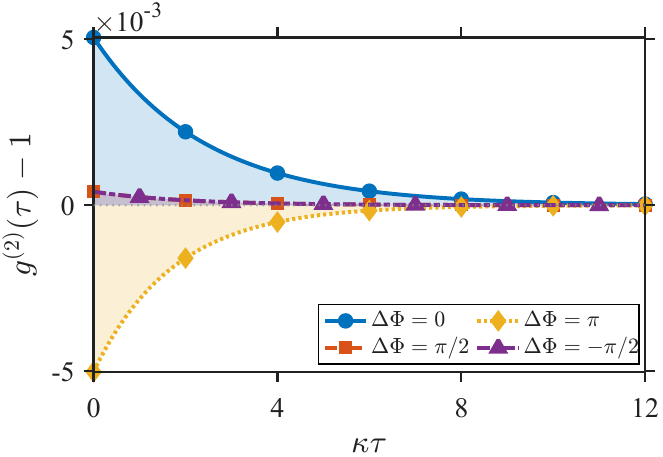}
\caption{
Post-click photon correlations $g^{(2)}(\tau)-1$ for representative feedback phases. 
The response is positive near $\Delta\Phi=0$, negative near $\Delta\Phi=\pi$, and strongly suppressed at $\Delta\Phi=\pm\pi/2$. 
The shaded areas indicate the integrated correlation response that determines the Mandel parameter.
}
\label{fig:counting_fluctuations}
\end{figure}

Figure~\ref{fig:counting_fluctuations} shows  \(g^{(2)}(\tau)-1\) for representative feedback phases.
Around \(\Delta\Phi=0\), the feedback enhances subsequent emission activity, yielding a positive correlation response and \(Q>0\). Around \(\Delta\Phi=\pi\), it suppresses subsequent emission activity, leading to a negative response and \(Q<0\).
Near  $\Delta\Phi=\pm\pi/2$, \(g^{(2)}(\tau)-1\) is small because the leading interference contribution between the stationary field and the feedback displacement is negligible.

The strong phase dependence of the output counting statistics contrasts with the weak variation of the normalized intracavity photon-number variance. The latter characterizes the cavity at a single time, whereas the Mandel parameter is sensitive to temporal correlations between emission events. The observed non-Poissonian statistics are therefore associated mainly with the phase-controlled temporal response generated by the feedback.

\subsection{Phase dependence and weak-feedback scaling}
\label{sec:weak_feedback_scaling}

\begin{figure*}[t]
	\center
	\includegraphics[width=1.95\columnwidth]{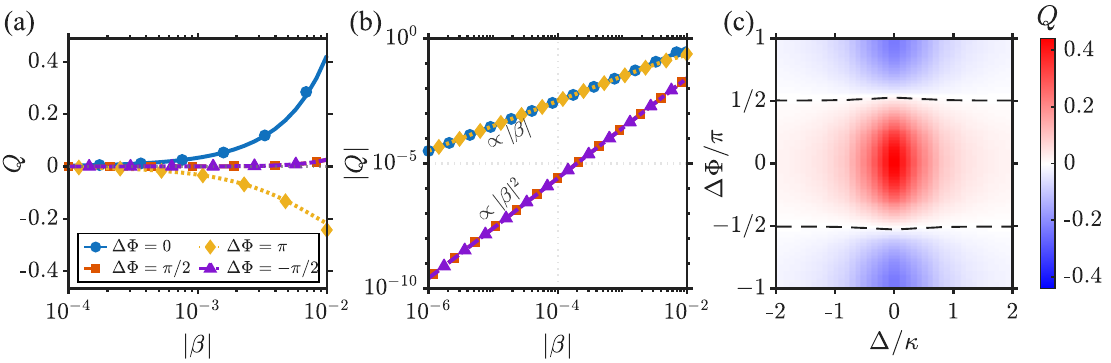}
	\caption{
	Phase dependence and weak-feedback scaling of the measurement noise.
	(a) Mandel parameter $Q$ as a function of the feedback amplitude $|\beta|$ for representative feedback phases.
	(b) Log--log plot of $|Q|$ versus $|\beta|$, showing linear scaling at generic phases and quadratic scaling at the specific phases $\Delta\Phi=\pm\pi/2$.
	Lines show the second-order analytical results, while markers denote the numerical full-counting-statistics results.
	(c) Mandel parameter $Q$ in the $(\Delta/\gamma,\Delta\Phi/\pi)$ plane at fixed feedback amplitude $|\beta|=10^{-2}$.
	The dashed curves mark the $Q=0$ boundaries separating the super- and sub-Poissonian counting regimes.
}
    \label{fig:mandel_scaling}
\end{figure*}

We consider the dependence of the photon-counting fluctuations on the feedback strength and phase.
Despite the non-linarity of the feedback operation, the Mandel parameter can be expressed in an surprisingly simple form,
\begin{equation}
    Q
    =
    2\nu+3\nu^2+O(|\beta|^3)
    \label{eq:Q_nu_weak}
\end{equation}
in the weak-feedback regime. The detailed derivation is given in Appendix~\ref{App:double_dyson}.
To second order in the feedback amplitude, $\nu$ is given by
\begin{equation}
\begin{aligned}
    \nu
    ={}&
    \frac{2\gamma|\Omega||\beta|}
    {(\gamma/2)^2+\Delta^2}
    \cos\Delta\Phi
    \\
    &+
    \left[
    1+
    \frac{2\gamma^2|\Omega|^2}
    {\left[(\gamma/2)^2+\Delta^2\right]^2}
    \right]
    |\beta|^2
    +
    O(|\beta|^3) \\
    \label{eq:nu_weak}
       ={}& \int_{0}^{\infty}dt \gamma\left [ n_{\rm fb,c} (t)- n_{\rm ss}\right].\\
\end{aligned}
\end{equation}
as shown in Appendix~\ref{app:mean_field}.
The first term arises from interference between the stationary cavity field and the feedback-induced displacement. 
It is linear in $|\beta|$, with its sign controlled by $\cos\Delta\Phi$. 
The quadratic term includes the direct displacement contribution and the change of the stationary state induced by the feedback. 
In the second equality, we have expressed this in terms of the mean occupation of the cavity  $ n_{\rm ss}$, and the mean cavity occupation in response to a single feedback action $n_{\rm fb,c}(t) = \left| \alpha_c(\tau)\right|^2$ , where $ \alpha_c(\tau) = \alpha_{\rm ss} + \beta e^{-(\gamma/2+i\Delta)\tau}$. In this form, one can interpret the parameter $\nu$ as the number of excess photon emitted in the response on a single feedback action.

Figure~\ref{fig:mandel_scaling}(a) shows the Mandel parameter as a function of the feedback amplitude for representative feedback phases. Around $\Delta\Phi=0$, the leading response is positive and gives super-Poissonian statistics, whereas around $\Delta\Phi=\pi$ it changes sign and produces sub-Poissonian statistics. 

Figure~\ref{fig:mandel_scaling}(b) resolves the asymptotic weak-feedback scaling.
For generic phases with $\cos\Delta\Phi\neq0$, the linear term remains finite, giving
\begin{equation}
    |Q|\propto|\beta|.
\end{equation}
At the quadrature phases $\Delta\Phi=\pm\pi/2$, the linear interference term vanishes, so that the quadratic contribution becomes leading and
\begin{equation}
    |Q|\propto|\beta|^2.
\end{equation}

The numerical curves approach slopes of one and two, respectively, consistent with the linear and quadratic asymptotic scaling. At finite feedback amplitudes, the $\Delta\Phi=0$ and $\pi$ branches in $|Q|$ are slightly separated by second-order corrections.

The phase-dependent sign control remains visible away from resonance. Figure~\ref{fig:mandel_scaling}(c) depicts $Q$ in the $(\Delta/\gamma,\Delta\Phi/\pi)$ plane for a fixed feedback amplitude.
The sign of \(Q\) remains strongly phase dependent, while detuning mainly modifies its magnitude through the cavity susceptibility.
At finite feedback strength, the $Q=0$ boundaries are shifted slightly away from the exact quadrature phases and approach $\Delta\Phi=\pm\pi/2$ in the weak-feedback limit.

\subsection{Higher-order counting cumulants}
\label{subsec:higher_order_cumulants}

The full-counting-statistics formalism in Eq.~\eqref{eq:tilted_liouvillian} provides direct access to higher-order cumulants. 
To determine whether the phase dependence identified from the Mandel parameter extends beyond the noise, we define
\begin{equation}
    Q_m
    =
    \frac{\kappa_m}{\kappa_1}
    -1,
    \qquad
    m\geq3.
    \label{eq:normalized_higher_cumulants}
\end{equation}
For Poissonian statistics, all long-time cumulant rates are equal to the mean counting rate, $\kappa_m=\kappa_1$, and therefore $Q_m=0$.

\begin{figure}[t]
	\center
	\includegraphics[width=0.95\columnwidth]{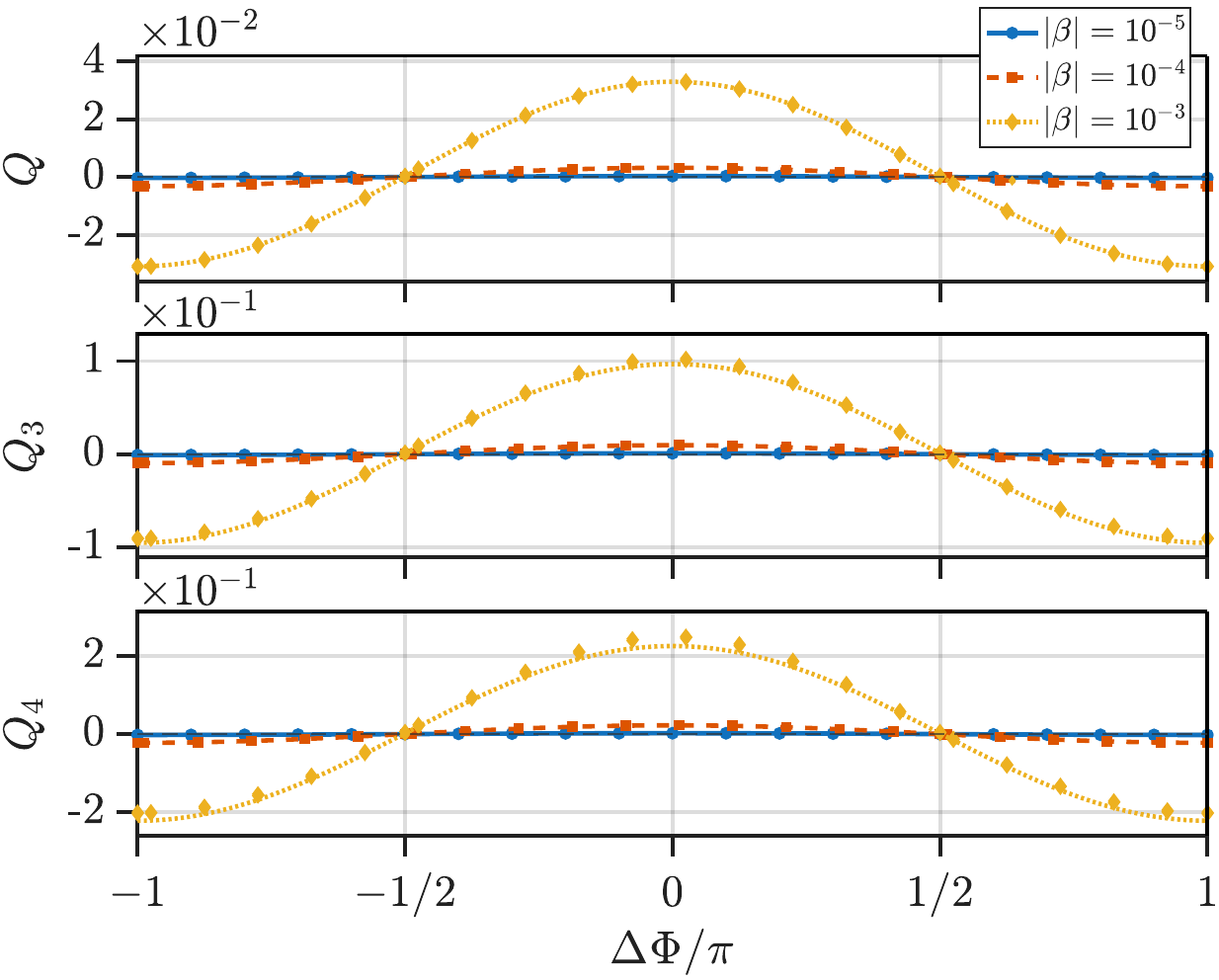}
	\caption{
Higher-order cumulants of the photon emission as a function of the feedback phase $\Delta\Phi$ for several feedback amplitudes.
The Mandel parameter $Q$ and the higher-order deviations $Q_3$ and $Q_4$ exhibit the same overall phase dependence.
Markers denote the numerical full-counting-statistics results, while lines show the weak-feedback analytical predictions in Eq.~\eqref{eq:higherOrderCumulants}.
}
    \label{fig:higher_cumulants}
\end{figure}

Figure~\ref{fig:higher_cumulants} compares $Q_3$ and $Q_4$ with the Mandel parameter $Q$ as a reference for several feedback amplitudes.  The nonzero \(Q_3\) and \(Q_4\) showcase that the feedback affects the counting statistics beyond the measurement noise.
The higher-order cumulants exhibit the same overall phase dependence as the noise. 
They are positive around $\Delta\Phi=0$, negative around $\Delta\Phi=\pi$, and strongly suppressed near $\Delta\Phi=\pm\pi/2$. 
At finite feedback strength, their zero crossings are shifted slightly away from exact $\Delta\Phi=\pm\pi/2$, consistent with the displacement of the $Q=0$ boundaries discussed above.

The origin of this common phase dependence can be understood from the weak-feedback expansion. 
The post-click mean-field dynamics determines the response parameter \(\nu\), which then enters the higher-order cumulants through the counting-field expansion. 
Within the primary-response approximation, the higher-order deviations are
\begin{equation}
Q_3\simeq6\nu,\qquad Q_4\simeq14\nu,
\label{eq:higherOrderCumulants}
\end{equation}
where \(\nu\) is the excess number of photons in response to a single feedback action  in Eq.~\eqref{eq:nu_weak}. The coefficients $6$ and $14$ arise from the counting-field structure of the leading order  correlations as shown in Appendix~\ref{app:higher_cumulants}. Thus, \(Q_3\) and \(Q_4\) inherit the phase dependence of \(\nu\), with different counting-field prefactors.
\section{Conclusion}
\label{Sec:Conclusion}

We have shown that click-triggered displacement feedback can generate and control non-Poissonian photon-counting statistics in a driven linear cavity. Rather than relying on intrinsic optical nonlinearities or nonclassical input states, the correlations arise from the conditional dynamics introduced by the measurement-feedback loop: each detected photon triggers a displacement that modifies the cavity state and hence the statistics of subsequent emissions.

Even when the displacement associated with each detection event is weak, its compounding action produces a significant modification of the stationary cavity field and photon occupation. The feedback phase determines the direction of this response and, more importantly, controls the temporal correlations in the emitted photon current. The integrated two-time correlation distinguishes super-Poissonian from sub-Poissonian counting, while a mean-field post-click response provides a simple description of the phase-sensitive mechanism underlying this behavior.

The weak-feedback regime reveals an additional sensitivity to the feedback phase. At generic phases, interference between the stationary cavity field and the feedback displacement produces a response linear in the feedback amplitude. At the quadrature phases, this linear contribution vanishes and the leading Mandel parameter response becomes quadratic. The same phase dependence is also observed in the higher-order  cumulants of the counting statistics, showing that the feedback modifies the emission statistics beyond the variance.

These results demonstrate that measurement-triggered feedback provides a route to controlling photon statistics in an otherwise linearly driven dissipative system. More generally, they show how conditional operations following individual quantum jumps can be used to engineer correlations in quantum trajectories without introducing intrinsic nonlinear interactions.

\section*{Acknowledgments}
G.E. acknowledges NSFC Grant No. W2432004. J.Y.L. acknowledges NSFC Grant No. 11774311. The authors used OpenAI ChatGPT (GPT-5.6 Sol) to assist with the derivation presented in Sec. Appendix C2. The physical model and assumptions were specified by the authors, and all AI-assisted material was independently checked and verified. The authors take full responsibility for the scientific content of this work.

\appendix

\section{Microscopic derivation of the feedback-resolved counting dynamics}
\label{app:microscopic_derivation}

In this Appendix, we derive the generalized master equation Eq.~\eqref{eq:tilted_liouvillian}
starting from the microscopic cavity--reservoir model introduced in the
main text. The derivation makes explicit  how the same measurement outcome both triggers the
feedback displacement and enters the full counting statistics. Throughout this construction, we assume a broadband Markovian reservoir, vacuum input in the monitored channel, unit detection
efficiency, negligible feedback delay, and an instantaneous
displacement following each registered photon.

\subsection{Photon counting of the outgoing temporal mode}

We perform photon-number detection on each outgoing temporal mode
$\hat b_t$. The measured observable is therefore
$\hat n_t=\hat b_t^\dagger\hat b_t$. In the continuous-time limit,
the probability for two or more photons to be emitted during a single
time interval is of order $dt^2$, so only the outcomes $m=0,1$ need to
be retained in leading order.

Conditioning on the photon-number outcome $m$ gives the cavity
measurement operator
\begin{equation}
\hat M_m
=
{}_t\langle m|
\hat V_t(dt)
|0\rangle_t .
\label{eq:app_Mm}
\end{equation}
where $\hat V_t(dt)$ is defined below Eq.~\eqref{eq:systemHamiltonian}.
This contraction removes the quantum degree of freedom of the photonic modes while retaining its measurement outcome $m$ as a classical record.

Expanding Eq.~\eqref{eq:app_Mm} to first order gives
\begin{equation}
\hat M_0
=
\hat I
-i\hat H_Sdt
-\frac{\gamma}{2}
\hat a^\dagger\hat a\,dt
+
O(dt^2),
\label{eq:app_M0}
\end{equation}
and
\begin{equation}
\hat M_1
=
\sqrt{\gamma dt}\,
\hat a
+
O(dt^{3/2}).
\label{eq:app_M1}
\end{equation}
The probability of detecting one photon is consequently
\begin{equation}
p_1
=
\operatorname{Tr}
\left[
\hat M_1\hat\rho\hat M_1^\dagger
\right]
=
\gamma
\operatorname{Tr}
\left[
\hat a^\dagger\hat a\hat\rho
\right]dt
+
O(dt^2).
\label{eq:app_click_probability}
\end{equation}
Hence a click corresponds microscopically to detecting a photon emitted
from the cavity into the monitored output channel.

\subsection{Click-triggered feedback}

A detected photon triggers the instantaneous intracavity displacement
\begin{equation}
\hat U
=
\hat D(\beta)
=
\exp\left(
\beta\hat a^\dagger
-
\beta^*\hat a
\right).
\label{eq:app_displacement}
\end{equation}
No feedback operation is applied when no photon is detected. The
measurement-conditioned cavity operation can therefore be written as
\begin{equation}
\hat K_m
=
\hat U^m\hat M_m,
\qquad m=0,1.
\label{eq:app_Km}
\end{equation}
In particular, for a detected photon,
\begin{equation}
\hat K_1
=
\sqrt{\gamma dt}\,
\hat U\hat a
+
O(dt^{3/2}).
\label{eq:app_K1}
\end{equation}
The ordering of the operators has a direct physical interpretation:
$\hat a$ describes emission into the monitored reservoir, whereas
$\hat U$ is subsequently applied conditional on detecting that emitted
photon. The feedback-dressed jump operator $\hat U\hat a$ therefore
represents the sequence
\emph{emission $\rightarrow$ detection $\rightarrow$ displacement}.

If the measurement record is disregarded after applying the feedback,
the unconditional cavity state evolves according to
\begin{equation}
\hat\rho(t+dt)
=
\sum_{m=0,1}
\hat K_m\hat\rho(t)\hat K_m^\dagger .
\label{eq:app_unconditional_map}
\end{equation}
Using Eqs.~\eqref{eq:app_M0}--\eqref{eq:app_K1} and taking
$dt\rightarrow0$ gives
\begin{equation}
\dot{\hat\rho}
=
-i[\hat H_S,\hat\rho]
+
\gamma
\left[
\hat U\hat a\hat\rho\hat a^\dagger\hat U^\dagger
-
\frac{1}{2}
\left\{
\hat a^\dagger\hat a,\hat\rho
\right\}
\right].
\label{eq:app_feedback_ME}
\end{equation}
Since $\hat U$ is unitary,
$(\hat U\hat a)^\dagger(\hat U\hat a)
=\hat a^\dagger\hat a$,this equation can be written as
\begin{equation}
\dot{\hat\rho}
=
-i[\hat H_S,\hat\rho]
+
\gamma
\mathcal D[\hat U\hat a]\hat\rho,
\label{eq:app_feedback_Lindblad}
\end{equation}
where
$\mathcal D[\hat A]\hat\rho
=\hat A\hat\rho\hat A^\dagger
-\frac{1}{2}\{\hat A^\dagger\hat A,\hat\rho\}$.

\subsection{Counting-field-resolved dynamics}

The same photon-detection record that controls the feedback also
defines the full counting statistics. Let $m_n=0,1$ denote the
measurement outcome at time $t$. The accumulated number up to time $\tau$
of detected photons is
\begin{equation}
N(\tau)
=
\sum_{t<\tau }m_t .
\label{eq:app_Nt}
\end{equation}
Thus, under the ideal detection assumptions used here, every detected
photon both increases $N(\tau)$ by one and triggers one application of
$\hat U$.

To retain the counting information before averaging over the
measurement outcomes, we associate the factor $e^{i\chi m}$ with each
outcome. The counting-field-resolved one-step evolution is therefore
\begin{equation}
\hat\rho_\chi(t+dt)
=
\sum_{m=0,1}
e^{i\chi m}
\hat U^m\hat M_m
\hat\rho_\chi(t)
\hat M_m^\dagger
\hat U^{\dagger m}.
\label{eq:app_counting_map}
\end{equation}
Equation~\eqref{eq:app_counting_map} makes explicit that the feedback
and the counting statistics are conditioned on the same microscopic
bath-measurement outcome.

Substituting Eqs.~\eqref{eq:app_M0} and \eqref{eq:app_M1} into
Eq.~\eqref{eq:app_counting_map} and retaining terms up to first order
in $dt$ gives
\begin{equation}
\dot{\hat\rho}_\chi
=
-i[\hat H_S,\hat\rho_\chi]
-\frac{\gamma}{2}
\left\{
\hat a^\dagger\hat a,\hat\rho_\chi
\right\}
+
\gamma e^{i\chi}
\hat U\hat a
\hat\rho_\chi
\hat a^\dagger\hat U^\dagger .
\label{eq:app_tilted_explicit}
\end{equation}
Introducing the physical feedback Liouvillian
$\mathcal L\hat\rho
=-i[\hat H_S,\hat\rho]
+\gamma\mathcal D[\hat U\hat a]\hat\rho$
and the jump superoperator
$\mathcal J\hat\rho
=\gamma\hat U\hat a\hat\rho\hat a^\dagger\hat U^\dagger$,
Eq.~\eqref{eq:app_tilted_explicit} becomes
\begin{equation}
\mathcal L_\chi
=
\mathcal L
+
\left(
e^{i\chi}-1
\right)
\mathcal J .
\label{eq:app_tilted_L}
\end{equation}

At $\chi=0$, the counting-field-resolved dynamics reduces to the
common quantum master equation including the feedback action. For finite $\chi$,
the moment-generating function given by $M(\chi,t)\equiv\operatorname{Tr}[\hat\rho_\chi(t)]$ generates the
statistics of photons detected in the monitored output channel~\cite{Landi2024}.

\section{Mean-field response}
\label{app:mean_field}

Taking the first moment of the feedback master equation gives
\begin{equation}
    \frac{d}{dt}\langle \hat a\rangle
    =
    -\left(\frac{\gamma}{2}+i\Delta\right)\langle\hat a\rangle
    -i\Omega
    +
    \gamma\beta
    \langle \hat a^\dagger\hat a\rangle .
    \label{eq:first_moment}
\end{equation}
Within the mean-field approximation,
$\langle \hat a^\dagger\hat a\rangle
\simeq
|\langle\hat a\rangle|^2$.
Denoting the stationary amplitude by
$\alpha_{\rm ss}=\langle\hat a\rangle_{\rm ss}$,
the stationary condition becomes
\begin{equation}
    0
    =
    -\left(\frac{\gamma}{2}+i\Delta\right)\alpha_{\rm ss}
    -i\Omega
    +
    \gamma\beta|\alpha_{\rm ss}|^2 .
    \label{eq:mf_stationary}
\end{equation}
In the absence of feedback, the stationary amplitude is
\begin{equation}
    \alpha_0
    =
    -\frac{i\Omega}
    {\gamma/2+i\Delta}.
    \label{eq:alpha0}
\end{equation}
Equation~\eqref{eq:mf_stationary} may therefore be written as
\begin{equation}
    \alpha_{\rm ss}
    =
    \alpha_0
    +
    \frac{\gamma\beta}
    {\gamma/2+i\Delta}
    |\alpha_{\rm ss}|^2 .
    \label{eq:mf_stationary_compact}
\end{equation}
For weak feedback,
\begin{equation}
    \alpha_{\rm ss}
    =
    \alpha_0
    +
    \frac{\gamma\beta}
    {\gamma/2+i\Delta}
    |\alpha_0|^2
    +
    O(|\beta|^2).
    \label{eq:mf_stationary_weak}
\end{equation}

We next characterize the response generated by a single detection-triggered feedback displacement.
Immediately after a detection event, the feedback changes the stationary cavity amplitude as
$\alpha_{\rm ss}\rightarrow\alpha_{\rm ss}+\beta$.
To isolate the response generated by this feedback action, subsequent detection events are neglected in the following mean-field construction.
The resulting cavity amplitude at time $\tau$ is
\begin{equation}
    \alpha_c(\tau)
    =
    \alpha_{\rm ss}
    +
    \beta
    e^{-(\gamma/2+i\Delta)\tau}.
    \label{eq:alpha_c}
\end{equation}
The corresponding mean-field emission rate is $   \gamma |\alpha_c(\tau)|^2$.
Using this emission rate, we define the excess number of photons in response to a single feedback action as
\begin{equation}
    \nu
    =
    \int_0^\infty
    \gamma \left[
    |\alpha_c(\tau)|^2
    -
    |\alpha_{\rm ss}|^2
    \right]d\tau,
    \label{eq:nu_definition}
\end{equation}
which is the same as in Eq.~\eqref{eq:nu_weak}. Substituting Eq.~\eqref{eq:alpha_c} gives
\begin{equation}
    \nu
    =
    2\gamma
    \operatorname{Re}
    \left[
    \frac{\alpha_{\rm ss}^{*}\beta}
    {\gamma/2+i\Delta}
    \right]
    +
    |\beta|^2 .
    \label{eq:nu_alpha}
\end{equation}

Expanding the stationary solution consistently in the weak-feedback regime yields
\begin{equation}
    \nu
    =
    \nu_1+\nu_2
    +
    O(|\beta|^3),
    \label{eq:nu_expand}
\end{equation}
with
\begin{equation}
    \nu_1
    =
    \frac{2\gamma|\Omega||\beta|}
    {(\gamma/2)^2+\Delta^2}
    \cos\Delta\Phi,
    \label{eq:nu1}
\end{equation}
and
\begin{equation}
    \nu_2
    =
    \left[
    1+
    \frac{2\gamma^2|\Omega|^2}
    {\left[(\gamma/2)^2+\Delta^2\right]^2}
    \right]
    |\beta|^2 .
    \label{eq:nu2}
\end{equation}
The linear contribution arises from interference between the stationary cavity field and the feedback-induced displacement, while the quadratic term contains the direct displacement contribution together with the correction associated with the feedback-induced shift of the stationary operating point. The mean-field construction therefore provides the primary-response parameter $\nu$ that enters the weak-feedback counting-field analysis below.

\section{Mandel parameter in the weak feedback regime}

\label{App:double_dyson}
In this appendix, we derive the expression for the Fano factor in the weak feedback regime in Eq.~\eqref{eq:Q_nu_weak}. In Appendix~\ref{app:formalExpansion}, we first derive a formal expansion expression. In Appendix~\ref{app:auxiliaryDensityMatrix}, we introduce an auxiliary density matrix formalism, which can used to straightforwardly evaluate the many terms of that expansion. In Appendix~\ref{app:evaluation}, we showcase the evaluation of the expansion terms. Subsequently, in Appendix~\ref{app:finalResult}, we reach the final result.

\subsection{Formal expansion}

\label{app:formalExpansion}

The Liouvillian operator in Eq.~\eqref{eq:tilted_liouvillian} can be distributed as
\begin{equation}
\mathcal L
=
\mathcal L_0+\mathcal V ,
\label{eq:liovillianDistribution}
\end{equation}
where $\mathcal L_0$ is the Liouvillian in the absence of feedback $\beta=0$, and 
\begin{equation}
\mathcal V
=
(\mathcal F_\beta-\mathbb I)\mathcal J_0 =\mathcal J -\mathcal J_0  ,
\end{equation}
is the perturbation caused by the feedback action. Thereby, $\mathcal J_0\hat\rho= \gamma\hat a\hat\rho\hat a^\dagger $ is the common jump operator and $\mathcal F_\beta \rho =\hat D(\beta)\rho \hat D^\dagger(\beta)$  is the shift operation.

Using these definitions, the Mandel parameter in Eq.~\eqref{eq:mandel_parameter} can be expressed as
\begin{equation}
Q
=\frac{2C}{\kappa_1},
\label{eq:fanoFactor}
\end{equation}
where we have introduced the current-current correlation function 
\begin{equation}
C
=
\int_0^\infty d\tau\,
{\rm Tr}
\left[
\mathcal J_0
e^{(\mathcal L_0+\mathcal V)\tau}
\Delta \mathcal J \hat\rho_{\rm ss}
\right].
\label{eq:jumpJumpCorrelationFunction}
\end{equation}
with the current fluctuation operator $\Delta \mathcal J =  \mathcal J -  \kappa_1 \mathbbm 1$ and the mean  current $\kappa_1={\rm Tr}\left[\mathcal J\hat\rho_{\rm ss}\right]$.

To obtain the Mandel parameter in second order in $\beta$, we must  expand $C$ and $\kappa_1$ to that same order. For brevity, we introduce the notation $\beta = \epsilon b$, with a complex valued $b$ and real-valued $\epsilon$, for which we carry out the perturbation expansion, i.e,
\begin{eqnarray}
\kappa_1
&=&
\kappa_1^{[0]}
+
\epsilon\kappa_1^{[1]}
+
\epsilon^2\kappa_1^{[2]}
+
O(\epsilon^3), \nonumber \\
C
&=&
\epsilon C^{[1]}
+
\epsilon^2C^{[2]}
+
O(\epsilon^3).
\end{eqnarray}
In the expansion for $C$ we have already used that $C^{[0]} =0$ as the counting statistics of the linear system is Poissonian.

To proceed, we introduce the following expansions
\begin{eqnarray}
	\mathcal V
	&=&
	\epsilon V^{[1]}
	+
	\epsilon^2V^{[2]}
	+
	O(\epsilon^3), \nonumber   \\
	\hat\rho_{\rm ss}
	&=&
	\hat\rho_0
	+
	\epsilon\hat\rho^{[1]}
	+
	\epsilon^2\hat\rho^{[2]}
	+
	O(\epsilon^3), \nonumber  \\
	\Delta \mathcal J &=&  \Delta \mathcal J^{[0]} + \epsilon \Delta \mathcal J^{[1]} + 	\epsilon^2 \Delta \mathcal J^{[2]} ,
	\label{eq:formalExpansion}
\end{eqnarray}
with  $\hat \rho_0 = \left| \alpha_0 \right> \left< \alpha_0 \right|$ for $\alpha_0 = i\Omega/(-i\omega -\gamma)$   being  the stationary state for $\beta =0$ [see Eq.~\eqref{eq:stationaryState}]. While the expansion of $\mathcal V$ can be directly obtained by derivation, the expansions of $\hat\rho_{\rm ss}$ and $\Delta \mathcal J $ (which implicitly depends on $\hat\rho_{\rm ss}  $  via $\kappa_1$) must be self-consistently obtained.  Using the formal expansion in Eq.~\eqref{eq:formalExpansion}, we thus find
\begin{eqnarray}
\kappa_1^{[0]}
&=&
{\rm Tr}
[
\mathcal J_0\hat\rho_0
],  \nonumber \\
\kappa_1^{[1]}
&=&
{\rm Tr}
[
\mathcal J_0\hat\rho^{[1]}
+
V^{[1]}\hat\rho_0
],  \nonumber  \\
\kappa_1^{[2]} &=& {\rm Tr}
[
\mathcal J_0\hat\rho^{[2]}
+
V^{[1]}\hat\rho^{[1]}
+
V^{[2]}\hat\rho_0
]. 
\label{eq:kappaExpansionTerms}
\end{eqnarray}
Moreover, the standard Dyson expansion of the time evolution operator in Eq.~\eqref{eq:jumpJumpCorrelationFunction} in orders of $\mathcal V$ yields 
\begin{eqnarray}
	C^{[1]}
	&=&
	-
	{\rm Tr}
	\left[
	\mathcal J_0
	\mathcal R_0
	\left(\Delta \mathcal J^{[1]}  \rho_0 + \Delta \mathcal J^{[0]}  \rho ^{[1]}\right) 
	\right]. \nonumber\\
	C^{[2]}
	&=& {}
	-
	{\rm Tr}
	\left[
	\mathcal J_0
	\mathcal R_0
	\left(\Delta \mathcal J^{[2]}  \rho_0 + \Delta \mathcal J^{[2]}  \rho ^{[1]} + \Delta \mathcal J^{[0]}  \rho ^{[2]} \right)  
	\right]
\nonumber	\\
	&&+
	{\rm Tr}
	\left[
	\mathcal J_0
	\mathcal R_0
	V^{[1]}
	\mathcal R_0
	\left(\Delta \mathcal J^{[1]}  \rho_0 +\Delta \mathcal J^{[0]}  \rho ^{[1]}\right) 
	\right],
\label{eq:correlationExpansionTerms}
\end{eqnarray}
which is expressed in terms of the resolvent of the unperturbed Liouvillian
\begin{equation}
\mathcal R_0
=
-\int_0^\infty dt\,
e^{\mathcal L_0t} Q_0.
\end{equation}
Thereby, we have introduce the projection $Q_0 \rho = \rho - \rho_0 {\rm Tr} \left[ \rho \right] $.
Crucially, the evaluation of Eqs.~\eqref{eq:kappaExpansionTerms} and \eqref{eq:correlationExpansionTerms}, requires the knowledge $\hat\rho^{[1]}$, $\hat\rho^{[2]}$. Using that
\begin{equation}
	\hat \rho_{ss} = \lim_{\tau \rightarrow \infty}e^{(\mathcal L_0+\mathcal V)\tau} \rho_0, 
	\label{eq:statinaryStateDefinition}
\end{equation}
a standard perturbation expansion of the time-evolution operator in Eq.~\eqref{eq:statinaryStateDefinition} in $\mathcal V$ similar as the one yielding Eq.~\eqref{eq:statinaryStateDefinition} shows that
\begin{eqnarray}
	\hat\rho^{[1]}
	&=&
	-\mathcal R_0
	V^{[1]}\hat\rho_0 , \nonumber \\
	\hat\rho^{[2]}
	&=&
	-\mathcal R_0
	\left(
	V^{[1]}\hat\rho^{[1]}
	+
	V^{[2]}\hat\rho_0
	\right).
\label{eq_app:corrections}
\end{eqnarray}
By inserting this into  Eqs.~\eqref{eq:kappaExpansionTerms} and \eqref{eq:correlationExpansionTerms}, we express both the first cumulant and the  current-current correlation function in terms of quantities of the linear system without feedback. Note that in doing so, the correlation function contains many terms, which must be evaluated. To enable a straightforward calculation of these terms, we introduce an auxiliary density matrix approach below. 

\subsection{Auxiliary density matrices}

\label{app:auxiliaryDensityMatrix}

As the Liouvillian in Eq.~\eqref{eq:liovillianDistribution} for $\beta=0$ reduces to an ordinary damped harmonic oscillator, the time evolution features the following property
\begin{eqnarray}
	e^{\mathcal L_0t}  \left| \alpha \right>  =   \left| \alpha_0 + (\alpha -\alpha_0)e^{zt} \right> 
	\label{eq:timeEvolutionCoherentState}
\end{eqnarray} 
with $z = -i \omega - \gamma$ and the stationary state $\left| \alpha_0\right>$ for $\alpha_0 = i\Omega/z$. Thus, an initial coherent state stays coherent for all times. To allow for a straightforward evaluation of the terms in Eq.~\eqref{eq:kappaExpansionTerms} and \eqref{eq:correlationExpansionTerms}, we define the following auxiliary matrices,
\begin{equation}
\chi_{mn}
=
\left.
\frac{\partial^{m+n}}
{\partial\alpha^m\partial(\alpha^*)^n}
\rho(\alpha)
\right|_{\alpha=\alpha_0},
\end{equation}
where $\rho(\alpha)  = \left|\alpha \right>\left< \alpha\right|$. Using relation Eq.~\eqref{eq:timeEvolutionCoherentState}, one can show that $\chi_{mn} $ has the following properties:

\begin{itemize}

	\item  \textit{Trace.} The trace of the auxiliary matrices fulfills
\begin{equation}
	{\rm Tr} \left( \chi_{mn} \right) =\delta_{m,0} \delta_{n,0}, 
	\label{eq:aux:trace}
\end{equation} 
which follows by deriving  ${\rm Tr} \left( \rho(\alpha) \right]=1$ with respect to $\alpha$ .
 
	\item  \textit{Time evolution.} Using Eq.~\eqref{eq:timeEvolutionCoherentState}, we find that the auxiliary density matrix evolves according to
	$
	e^{\mathcal L_0t}\chi_{mn}
	=
	e^{-(mz+nz^*)t}\chi_{mn}
	$ 
	, from which we obtain
	\begin{equation}
	R_0\chi_{mn}
	=
	-\frac{1}{mz+nz^*}\chi_{mn},
	\label{eq:aux:timeEvolution}
	\end{equation}
	for $m+n>0$. Otherwise, we trivially have that $R_0\chi_{00} =0 $.
	
	\item  \textit{Jump operator}. The jump operator acts as follows on the auxiliary density matrix
\begin{eqnarray}
J_0\chi_{mn}
&=&
\kappa_1^{[0]}\chi_{mn}
+
m\gamma\alpha_0^*\chi_{m-1,n} \nonumber \\
&+&
n\gamma\alpha_0\chi_{m,n-1}
+
mn\gamma\chi_{m-1,n-1},
\label{eq:aux:jumpOperator}
\end{eqnarray}
with $\kappa_1^{[0]} ={\rm Tr} [\mathcal J_0\hat\rho_0]$, which can be shown by deriving $J_0\rho(\alpha)=\gamma|\alpha|^2\rho(\alpha),$ with respect to $\alpha$. 

	\item  \textit{Feedback action.} Representing $\mathcal F_\beta \rho =\hat D(\beta)\rho \hat D^\dagger(\beta)$ as $\mathcal F_\beta = e^{\epsilon \mathcal K} $ with $\mathcal K  \rho = [b\hat a^\dagger-b^*\hat a,X]  $, the auxiliary density matrices fulfill
	\begin{equation}
	\mathcal K\chi_{mn}
	=
	b\chi_{m+1,n}
	+
	b^*\chi_{m,n+1},
	\label{eq:aux:feedback}
	\end{equation}
	which can be verified by deriving $\mathcal F_\beta \rho =  \left|\alpha_0 + \epsilon b \right> \left<\alpha_0+ \epsilon b  \right|$ several times with respect to $\alpha =\alpha_0+ \epsilon b $ (and $\alpha^*$) and one time with respect to $\epsilon$ at $\epsilon=0$.

\end{itemize}

\subsection{Evaluation of the correlation functions}

\label{app:evaluation}

Equipped with the properties of the auxiliary density matrix in Appendix~\ref{app:auxiliaryDensityMatrix}, it is straightforward to evaluate the correlation functions in Eq.~\eqref{eq:kappaExpansionTerms} and \eqref{eq:correlationExpansionTerms}.

We trivially find $\kappa_1^{[0]} =\gamma|\alpha_0|^2$. To evaluate the next term in the expansion, we first recognize that
\begin{eqnarray}
	\rho^{[1]}
	&=&
	-R_0V^{[1]}\rho_0
	\nonumber\\
	&=&
	\kappa_1^{[0]}
	\left(
	\frac{b}{z}\chi_{10}
	+
	\frac{b^*}{z^*}\chi_{01}
	\right)
	\label{eq_app:rho_1}
\end{eqnarray}
according to Eqs.~\eqref{eq:aux:timeEvolution}, \eqref{eq:aux:jumpOperator}, and \eqref{eq:aux:feedback}. For this reason,
\begin{eqnarray}
	\kappa_1^{[1]}
	&=&
	\operatorname{Tr}[J_0\rho^{[1]}] \nonumber
	\\
	&=&
	\kappa_1^{[0]}\gamma
	\left(
	\frac{\alpha_0^*b}{z}
	+
	\frac{\alpha_0b^*}{z^*}
	\right). \nonumber  \\
	&\equiv & \kappa_1^{[0]}\nu^{[1]},
	\label{eq:firstCumulant:orderOne}
\end{eqnarray}
where we have used that $\operatorname{Tr}[V^{[1]}X] = 0$ (as $V^{[1]}$ contains a commutator operation), as well as Eqs.~\eqref{eq:aux:trace} and \eqref{eq:aux:jumpOperator}.  The other terms in Eq.~\eqref{eq:kappaExpansionTerms} and \eqref{eq:correlationExpansionTerms} can be evaluated similarily. Doing so yields
\begin{equation}
	\kappa_1^{[2]}
	=
	\kappa_1^{[0]}
	\left[
	(\nu^{[1]})^2
	+
	\frac{\gamma|b|^2}{z+z^*}
	+
	\frac{
		\gamma \kappa_1^{[0]}|b|^2
	}{|z|^2}
	\right],
	\label{eq:firstCumulant:orderTwo}
\end{equation}
as well as
\begin{eqnarray}
C^{[1]}
&=&
\kappa_1^{[0]}\gamma
\left(
\frac{\alpha_0^*b}{z}
+
\frac{\alpha_0b^*}{z^*}
\right),  \nonumber \\
	C^{[2]}
	&=&
	\kappa_1^{[0]}
	\left[
	\frac52(\nu^{[1]})^2
	+
	\frac{\gamma|b|^2}{z+z^*}
	+
	\frac{
		2\gamma \kappa_1^{[0]}|b|^2
	}{|z|^2}
	\right],
	\label{eq:correlation:orderOneTwo}
\end{eqnarray}
which are the desired expansion coefficients for the first current cumulant and the current-current correlation functions.

\subsection{Mandel parameter  in second-order perturbation}

\label{app:finalResult}

Inserting Eqs.~\eqref{eq:firstCumulant:orderOne}- \eqref{eq:correlation:orderOneTwo} into Eq.~\eqref{eq:fanoFactor}, and expanding the fraction up to second order in $\epsilon$, we finally obtain
\begin{equation}
	Q
	=
	2\epsilon\nu^{[1]}
	+
	\epsilon^2
	\left[
	2\nu^{[2]}
	+
	3(\nu^{[1]})^2
	\right]
	+
	O(\epsilon^3).
\end{equation}
which is expressed in terms of 
\begin{equation}
	\nu^{[2]}
	=
	|b|^2
	+
	\frac{
		2\gamma\kappa_1^{[0]}|b|^2
	}{
		(\gamma/2)^2+\Delta^2
	}.
\end{equation}
Interestingly, this can be brought into the form
\begin{equation}
	Q = 2\nu+3\nu^2+O(\epsilon^3)
\end{equation}
where
\begin{equation}
	\nu
	=
	2\gamma\,{\rm Re}
	\left[
	\frac{\alpha_{\rm ss}^{*}\beta}
	{\gamma/2+i\Delta}
	\right]
	+
	|\beta|^2 .
\end{equation}
is the surplus of emitted photons upon a single feedback action expressed in terms of $\alpha_{\rm ss}$ given in Eq.~\eqref{eq:meanField}.

\section{Higher-order  cumulants}
\label{app:higher_cumulants}

Here we identify the leading contribution of the  higher-order cumulants.
Introducing $s=i\chi$ and the auxiliary variable $z=e^s-1$, we first expand the long-time scaled cumulant-generating function as
\begin{equation}
    \theta(s) = \lim_{t \rightarrow \infty}\frac{K(-is,t)}{t} 
    =
    \sum_{r=1}^{\infty}
    \frac{C_r}{r!}z^r ,
    \label{eq:scgf_z}
\end{equation}
which is called the factorial cumulant-generating function when regarded as a function of $z$~\cite{Kambly2011, Mandel1995,Beenakker2001}. The $C_r$ , commonly referred to as indices of dispersion,  indicate how close a probability distribution is to a Poissonian one.
Substituting $z=e^s-1$ and re-expanding Eq.~\eqref{eq:scgf_z} in powers of $s$, we compare it with the ordinary cumulant expansion
\begin{equation}
    \theta(s)
    =
    \sum_{m=1}^{\infty}
    \frac{\kappa_m}{m!}s^m .
    \label{eq:scgf_s}
\end{equation}
Matching equal powers of $s$ gives, in particular,
\begin{equation}
    C_1=\kappa_1,
    \qquad
    C_2
    =
    \kappa_2-\kappa_1
    =
    \kappa_1 Q .
    \label{eq:C2_Q}
\end{equation}

Using the leading weak-feedback result
$Q=2\nu+O(\nu^2)$ obtained from the counting-field expansion above, Eq.~\eqref{eq:C2_Q} gives
\begin{equation}
    C_2
    =
    2\kappa_1\nu
    +
    O(\nu^2).
    \label{eq:C2_nu}
\end{equation}
The same coefficient $C_2$ contributes to higher-order cumulants through the term
$C_2(e^s-1)^2/2$ in Eq.~\eqref{eq:scgf_z}.
Since
\begin{equation}
    \left.
    \frac{d^m}{ds^m}
    (e^s-1)^2
    \right|_{s=0}
    =
    2^m-2 ,
    \label{eq:z2_derivative}
\end{equation}
taking the $m$th derivative and normalizing by $\kappa_1$ directly gives
\begin{equation}
    Q_m
    =
    (2^m-2)\nu
    +
    O(\nu^2) .
    \label{eq:Qm_general}
\end{equation}

For the third and fourth cumulants considered in the main text,
\begin{equation}
    Q_3
    =
    6\nu
    +
    O(\nu^2),
    \qquad
    Q_4
    =
    14\nu
    +
    O(\nu^2).
    \label{eq:Q34}
\end{equation}
Thus, the leading phase and feedback-strength dependence of the higher-order  cumulants is governed by the same response parameter $\nu$ that controls the Mandel parameter, while the factor $2^m-2$ determines the relative magnitude at different cumulant orders.


\bibliography{bibliography}

\end{document}